\documentclass[times, twocolumn]{aastex63}
\usepackage{xspace}
\usepackage{amsmath}
\usepackage{graphicx}
\usepackage{bm}
\usepackage{tikz}
\let\tablenum\relax
\usepackage{siunitx}
\usepackage{xcolor}
\newcommand{\changes}[1]{\text{#1}}

\newcommand{\tess}{\emph{TESS}\xspace}

\newcommand{\tres}{TRES\xspace}
\newcommand{\kelt}{KELT\xspace}
\newcommand{\neid}{NEID\xspace}
\newcommand{\alopeke}{\`{}Alopeke\xspace}
\newcommand{\pharo}{PHARO\xspace}

\newcommand{\ut}{UT\xspace}

\newcommand{\tic}{TIC-142394656\xspace}

\newcommand{\toi}{TOI-1268\xspace}
\newcommand{\toib}{TOI-1268b\xspace}

\providecommand{\msun}{\ensuremath{M_\Sun}}
\providecommand{\rsun}{\ensuremath{R_\Sun}}

\providecommand{\rj}{\ensuremath{R_{\rm Jup}}}

\newcommand{\mstar}{$0.90\pm0.13$}
\newcommand{\rstar}{$0.86\pm0.02$}
\newcommand{\rhostar}{$1.98\pm0.33$}
\newcommand{\logg}{$4.52\pm0.07$}
\newcommand{\teff}{$5257\pm40$}
\newcommand{\feh}{$+0.17\pm0.06$}

\newcommand{\per}{8.157728$^{+0.000005}_{-0.000005}$}
\newcommand{\midt}{2458703.5895$^{+0.0003}_{-0.0002}$}
\newcommand{\rprs}{0.089$^{+0.001}_{-0.001}$}
\newcommand{\rp}{0.747$^{+0.018}_{-0.018}$}

\newcommand{\DTrhocirc}{1.450$^{+0.033}_{-0.049}$}
\newcommand{\DTb}{0.135$^{+0.065}_{-0.049}$}

\newcommand{\DTvsini}{4.183$^{+0.306}_{-0.231}$}
\newcommand{\DTlam}{39.997$^{+7.223}_{-9.904}$}
\newcommand{\DTua}{0.239$^{+0.071}_{-0.074}$}
\newcommand{\DTub}{0.478$^{+0.151}_{-0.137}$}
\newcommand{\DTnonrotv}{2.020$^{+0.134}_{-0.131}$}

\newcommand{\DTincl}{89.552$^{+0.165}_{-0.222}$}
\newcommand{\DTaor}{17.215$^{+0.131}_{-0.196}$}
\newcommand{\DTap}{0.0688$^{+0.0017}_{-0.0018}$}

\newcommand{\SERVALrhocirc}{1.407$^{+0.064}_{-0.094}$}
\newcommand{\SERVALb}{0.246$^{+0.121}_{-0.128}$}

\newcommand{\SERVALvsini}{4.303$^{+0.553}_{-0.454}$}
\newcommand{\SERVALlam}{24.873$^{+12.963}_{-13.134}$}
\newcommand{\SERVALua}{0.236$^{+0.076}_{-0.077}$}
\newcommand{\SERVALub}{0.462$^{+0.154}_{-0.146}$}
\newcommand{\SERVALjitter}{2.255$^{+1.861}_{-1.683}$}

\newcommand{\SERVALincl}{89.318$^{+0.354}_{-0.337}$}
\newcommand{\SERVALaor}{17.043$^{+0.256}_{-0.387}$}
\newcommand{\SERVALap}{0.0680$^{+0.0020}_{-0.0021}$}

\newcommand{\DRPrhocirc}{1.437$^{+0.051}_{-0.099}$}
\newcommand{\DRPb}{0.191$^{+0.149}_{-0.130}$}

\newcommand{\DRPvsini}{4.472$^{+0.454}_{-0.392}$}
\newcommand{\DRPlam}{13.623$^{+14.145}_{-9.572}$}
\newcommand{\DRPua}{0.235$^{+0.072}_{-0.074}$}
\newcommand{\DRPub}{0.473$^{+0.150}_{-0.142}$}
\newcommand{\DRPjitter}{3.473$^{+1.791}_{-2.243}$}

\newcommand{\DRPincl}{89.469$^{+0.361}_{-0.413}$}
\newcommand{\DRPaor}{17.164$^{+0.199}_{-0.403}$}
\newcommand{\DRPap}{0.0684$^{+0.0019}_{-0.0021}$}

\newcommand{\SERVALecc}{0.13$^{+0.27}_{-0.13}$}
\newcommand{\SERVALomg}{212$^{+25}_{-20}, 328^{+19}_{-26}$}

\newcommand{\DRPecc}{0.13$^{+0.24}_{-0.13}$}
\newcommand{\DRPomg}{210$^{+25}_{-19}, 329^{+21}_{-25}$}

\newcommand{\DTecc}{0.12$^{+0.24}_{-0.12}$}
\newcommand{\DTomg}{210$^{+24}_{-19}, 329^{+20}_{-27}$}

\received{December 10, 2021}
\revised{January 19, 2022}
\accepted{January 20, 2022}
\submitjournal{AAS Journals}

\shorttitle{TOI-1268b: A Young Warm Saturn Aligned with Its Cool Host Star}
\shortauthors{Dong, Huang, Zhou et al.}

\begin{document}

\title{NEID Rossiter-McLaughlin Measurement of TOI-1268b: A Young Warm Saturn Aligned with Its Cool Host Star}

\newcommand{\PSUAA}{Department of Astronomy \& Astrophysics, 525 Davey Laboratory, The Pennsylvania State University, University Park, PA, 16802, USA}
\newcommand{\PSUCEHW}{Center for Exoplanets and Habitable Worlds, 525 Davey Laboratory, The Pennsylvania State University, University Park, PA, 16802, USA}
\newcommand{\USQAstro}{University of Southern Queensland, Centre for Astrophysics, West Street, Toowoomba, QLD 4350 Australia}
\newcommand{\CfA}{Center for Astrophysics \textbar \ Harvard \& Smithsonian, 60 Garden Street, Cambridge, MA 02138, USA}
\newcommand{\MSUAstro}{Department of Physics and Astronomy, Michigan State University, East Lansing, MI 48824, USA}
\newcommand{\Birmingham}{School of Physics \& Astronomy, University of Birmingham, Edgbaston, Birmingham B15 2TT, United Kingdom}
\newcommand{\KavliMIT}{Department of Physics and Kavli Institute for Astrophysics and Space Research, Massachusetts Institute of Technology, Cambridge, MA 02139, USA}
\newcommand{\PSETI}{Penn State Extraterrestrial Intelligence Center, 525 Davey Laboratory, The Pennsylvania State University, University Park, PA, 16802, USA}
\newcommand{\UA}{Steward Observatory, The University of Arizona, 933 N.\ Cherry Ave, Tucson, AZ 85721, USA}
\newcommand{\UAA}{Department of Astronomy and Steward Observatory, University of Arizona, Tucson, AZ 85721, USA}
\newcommand{\Penn}{Department of Physics and Astronomy, University of Pennsylvania, 209 S 33rd St, Philadelphia, PA 19104, USA}
\newcommand{\Caltech}{Department of Astronomy, California Institute of Technology, Pasadena, CA 91125, USA}
\newcommand{\STScI}{Space Telescope Science Institute, 3700 San Martin Dr, Baltimore, MD 21218, USA}
\newcommand{\JHU}{Department of Physics and Astronomy, Johns Hopkins University, 3400 N Charles St, Baltimore, MD 21218, USA}
\newcommand{\GoddardESAL}{Exoplanets and Stellar Astrophysics Laboratory, NASA Goddard Space Flight Center, Greenbelt, MD 20771, USA}
\newcommand{\GoddardISTD}{Instrument Systems and Technology Division, NASA Goddard Space Flight Center, Greenbelt, MD 20771, USA}
\newcommand{\GSFC}{NASA Goddard Space Flight Center, Greenbelt, MD 20771, USA}
\newcommand{\NOAO}{NSF's National Optical-Infrared Astronomy Research Laboratory, 950 N.\ Cherry Ave., Tucson, AZ 85719, USA}
\newcommand{\Macquarie}{Department of Physics and Astronomy, Macquarie University, Balaclava Road, North Ryde, NSW 2109, Australia}
\newcommand{\NIST}{National Institute of Standards \& Technology, 325 Broadway, Boulder, CO 80305, USA}
\newcommand{\CUBoulder}{Department of Physics, 390 UCB, University of Colorado, Boulder, CO 80309, USA}
\newcommand{\JPL}{Jet Propulsion Laboratory, California Institute of Technology, 4800 Oak Grove Drive, Pasadena, California 91109}
\newcommand{\UCI}{Department of Physics \& Astronomy, The University of California, Irvine, Irvine, CA 92697, USA}
\newcommand{\Carleton}{Carleton College, One North College St., Northfield, MN 55057, USA}
\newcommand{\PSUICS}{Institute for Computational and Data Sciences, The Pennsylvania State University, University Park, PA, 16802, USA}
\newcommand{\PSUCASt}{Center for Astrostatistics, 525 Davey Laboratory, The Pennsylvania State University, University Park, PA, 16802, USA}
\newcommand{\NESSF}{NASA Earth and Space Science Fellow}
\newcommand{\Princeton}{Department of Astrophysical Sciences, Princeton University, 4 Ivy Lane, Princeton, NJ 08540, USA}
\newcommand{\RUSSELL}{Henry Norris Russell Fellow}
\newcommand{\IAS}{Institute for Advance Study, 1 Einstein Drive, Princeton, NJ 08540, USA}
\newcommand{\Tsinghua}{Department of Astronomy, Tsinghua University, Beijing 100084, China}
\newcommand{\Lafayette}{Department of Physics, Lafayette College, 730 High St., Easton, PA 18042, USA}
\newcommand{\NASAAmes}{NASA Ames Research Center, Moffett Field, CA 94035, USA}
\newcommand{\MITEaps}{Department of Earth, Atmospheric, and Planetary Sciences, Massachusetts Institute of Technology, Cambridge, MA 02139, USA}
\newcommand{\MITAero}{Department of Aeronautics and Astronautics, Massachusetts Institute of Technology, Cambridge, MA 02139, USA}
\newcommand{\SAI}{Sternberg Astronomical Institute, M.V. Lomonosov Moscow State University, 13, Universitetskij pr., 119234, Moscow, Russia}
\newcommand{\NExScI}{NASA Exoplanet Science Institute, Caltech/IPAC, Mail Code 100-22, 1200 E. California Blvd., Pasadena, CA 91125, USA}
\newcommand{\Vanderbilt}{Department of Physics and Astronomy, Vanderbilt University, Nashville, TN 37235, USA}
\newcommand{\TexasAustin}{Department of Astronomy, The University of Texas at Austin, Austin, TX 78712, USA}
\newcommand{\Kotizarovci}{Kotizarovci Observatory, Sarsoni 90, 51216 Viskovo, Croatia}
\newcommand{\StephenAustin}{Department of Physics, Engineering and Astronomy, Stephen F. Austin State University, 1936 North St, Nacogdoches, TX 75962, USA}
\newcommand{\Oukaimeden}{Oukaimeden Observatory, High Energy Physics and Astrophysics Laboratory, Cadi Ayyad University, Marrakech, Morocco}
\newcommand{\MauryLewin}{The Maury Lewin Astronomical Observatory, Glendora,California.91741. USA}
\newcommand{\ValenciaAstro}{Departamento de Astronom\'{\i}a y Astrof\'{\i}sica, Universidad de Valencia, E-46100 Burjassot, Valencia, Spain}
\newcommand{\ValenciaObs}{Observatorio Astron\'omico, Universidad de Valencia, E-46980 Paterna, Valencia, Spain}
\newcommand{\Wellesley}{Department of Astronomy, Wellesley College, Wellesley, MA 02481, USA}
\newcommand{\Calou}{Observatori de Ca l'Ou, Carrer de dalt 18, Sant Martí Sesgueioles 08282, Barcelona, Spain}
\newcommand{\IACSpain}{Instituto de Astrof\'isica de Canarias (IAC), E-38205 La Laguna, Tenerife, Spain}
\newcommand{\ULLSpain}{Departamento de Astrof\'isica, Universidad de La Laguna (ULL), E-38206 La Laguna, Tenerife, Spain}
\newcommand{\ARU}{Astrobiology Research Unit, Universit\'e de Li\`ege, 19C All\'ee du 6 Ao\^ut, 4000 Li\`ege, Belgium}
\newcommand{\STAR}{Space sciences, Technologies and Astrophysics Research (STAR) Institute, Universit\'e de Li\`ege, Belgium}
\newcommand{\BerkeleyAstro}{Department of Astronomy, University of California, Berkeley, Berkeley, CA, USA}
\newcommand{\MarylandAstro}{Department of Astronomy, University of Maryland, College Park, College Park, MD, USA}
\newcommand{\Dartmouth}{Department of Physics and Astronomy, Dartmouth College, Hanover, NH 03755, USA}

\correspondingauthor{Jiayin Dong}
\email{jdong@psu.edu}

\author[0000-0002-3610-6953]{Jiayin Dong}
\affil{\PSUAA}
\affil{\PSUCEHW}

\author[0000-0003-0918-7484]{Chelsea X. Huang} 
\affil{\USQAstro}
\affil{\KavliMIT}

\author[0000-0002-4891-3517]{George Zhou} 
\affil{\USQAstro}

\author[0000-0001-9677-1296]{Rebekah I. Dawson} 
\affil{\PSUAA}
\affil{\PSUCEHW}
\affil{\PSUICS}
\affil{\PSUCASt}

\author[0000-0001-7409-5688]{Gudmundur K. Stefánsson} 
\altaffiliation{\RUSSELL}
\affil{\Princeton}

\author[0000-0003-4384-7220]{Chad F.\ Bender} 
\affil{\UA}

\author[0000-0002-6096-1749]{Cullen H.\ Blake} 
\affil{\Penn}

\author[0000-0001-6545-639X]{Eric B.\ Ford} 
\affil{\PSUAA}
\affil{\PSUCEHW}
\affil{\PSUICS}
\affil{\PSUCASt}

\author[0000-0003-1312-9391]{Samuel Halverson} 
\altaffiliation{Sagan Fellow}
\affil{\JPL}
\affil{\KavliMIT}

\author[0000-0001-8401-4300]{Shubham Kanodia} 
\affil{\PSUAA}
\affil{\PSUCEHW}
\affil{\PSETI}

\author[0000-0001-9596-7983]{Suvrath Mahadevan} 
\altaffiliation{NEID Principal Investigator}
\affil{\PSUAA}
\affil{\PSUCEHW}

\author[0000-0003-0241-8956]{Michael W.\ McElwain} 
\affil{\GoddardESAL} 

\author[0000-0001-8720-5612]{Joe P.\ Ninan} 
\affil{\PSUAA}
\affil{\PSUCEHW}

\author[0000-0003-0149-9678]{Paul Robertson} 
\altaffiliation{NEID Instrument Team Project Scientist}
\affil{\UCI}

\author[0000-0001-8127-5775]{Arpita Roy} 
\affil{\STScI}
\affil{\JHU}

\author[0000-0002-4046-987X]{Christian Schwab} 
\affil{\Macquarie}

\author[0000-0002-5951-8328]{Daniel J. Stevens} 
\affil{\PSUAA}
\affil{\PSUCEHW}

\author[0000-0002-4788-8858]{Ryan C. Terrien} 
\affil{\Carleton}

\author[0000-0001-7246-5438]{Andrew Vanderburg} 
\affil{\KavliMIT}

\author[0000-0001-9811-568X]{Adam L. Kraus} 
\affil{\TexasAustin}

\author[0000-0001-7371-2832]{Stephanie Douglas} 
\affil{\Lafayette}

\author[0000-0003-4150-841X]{Elisabeth Newton} 
\affil{\Dartmouth}

\author[0000-0001-7337-5936]{Rayna Rampalli} 
\affil{\Dartmouth}

\author[0000-0001-9626-0613]{Daniel M. Krolikowski} 
\affil{\TexasAustin}

\author[0000-0001-6588-9574]{Karen A.\ Collins} 
\affil{\CfA}

\author[0000-0001-8812-0565]{Joseph E. Rodriguez} 
\affil{\MSUAstro}

\author[0000-0002-2457-7889]{Dax L. Feliz} 
\affil{\Vanderbilt}

\author{Gregor Srdoc} 
\affil{\Kotizarovci}

\author[0000-0002-0619-7639]{Carl Ziegler} 
\affil{\StephenAustin}

\author[0000-0003-1464-9276]{Khalid Barkaoui} 
\affil{\ARU}
\affiliation{\MITEaps}

\author[0000-0003-1572-7707]{Francisco J. Pozuelos} 
\affil{\ARU}
\affil{\STAR}

\author[0000-0001-8923-488X]{Emmanuel Jehin} 
\affil{\STAR}

\author[0000-0003-1462-7739]{Micha\"{e}l c} 
\affil{\ARU}

\author[0000-0001-6285-9847]{Zouhair Benkhaldoun} 
\affil{\Oukaimeden}

\author[0000-0003-0828-6368]{Pablo Lewin} 
\affiliation{\MauryLewin}

\author[0000-0002-6482-2180]{Raquel For\'{e}s-Toribio} 
\affil{\ValenciaAstro}
\affil{\ValenciaObs}

\author[0000-0001-9833-2959]{Jose A. Mu\~noz} 
\affil{\ValenciaAstro}
\affil{\ValenciaObs}

\author[0000-0001-9504-1486]{Kim K. McLeod} 
\affil{\Wellesley}

\author{Fiona Powers \"Ozyurt} 
\affil{\Wellesley}

\author[0000-0001-9927-7269]{Ferran Grau Horta} 
\affiliation{\Calou}

\author[0000-0001-9087-1245]{Felipe Murgas} 
\affiliation{\IACSpain}
\affiliation{\ULLSpain}

\author[0000-0001-9911-7388]{David W. Latham} 
\affil{\CfA}

\author[0000-0002-8964-8377]{Samuel N. Quinn} 
\affil{\CfA}

\author[0000-0001-6637-5401]{Allyson Bieryla} 
\affil{\CfA}

\author[0000-0002-2532-2853]{Steve~B.~Howell} 
\affil{\NASAAmes}

\author[0000-0003-2519-6161]{Crystal~L.~Gnilka} 
\affil{\NExScI}
\affil{\NASAAmes}

\author[0000-0002-5741-3047]{David R. Ciardi} 
\affil{\NExScI}

\author[0000-0003-2527-1598]{Michael B. Lund} 
\affil{\NExScI}

\author[0000-0001-8189-0233]{Courtney D. Dressing} 
\affil{\BerkeleyAstro}

\author[0000-0002-8965-3969]{Steven Giacalone} 
\affil{\BerkeleyAstro}

\author[0000-0002-2454-768X]{Arjun B. Savel} 
\affil{\MarylandAstro}

\author[0000-0003-0647-6133]{Ivan A. Strakhov} 
\affil{\SAI}

\author[0000-0003-3469-0989]{Alexander A.\ Belinski} 
\affiliation{\SAI}

\author[0000-0003-2058-6662]{George R. Ricker} 
\affil{\KavliMIT}

\author[0000-0002-6892-6948]{S. Seager} 
\affil{\MITEaps}
\affil{\KavliMIT}
\affil{\MITAero}

\author[0000-0002-4265-047X]{Joshua N.\ Winn} 
\affil{\Princeton}

\author[0000-0002-4715-9460]{Jon M. Jenkins} 
\affil{\NASAAmes}

\author[0000-0002-5286-0251]{Guillermo Torres} 
\affil{\CfA}

\author[0000-0001-8120-7457]{Martin Paegert} 
\affil{\CfA}

\begin{abstract}
Close-in gas giants present a surprising range of stellar obliquity, the angle between a planet's orbital axis and its host star's spin axis. It is unclear whether the obliquities reflect the planets' dynamical history (e.g., aligned for in situ formation or disk migration versus misaligned for high-eccentricity tidal migration) or whether other mechanisms (e.g., primordial misalignment or planet-star interactions) are more important in sculpting the obliquity distribution.
Here we present the stellar obliquity measurement of \toi (\tic, $V_{\rm mag} {\sim} 10.9$), a young K-type dwarf hosting an 8.2-day period, Saturn-sized planet. \toi's lithium abundance and rotation period suggest the system age between the ages of Pleiades cluster (${\sim}120$ Myr) and Prasepe cluster (${\sim}670$ Myr). 
Using the newly commissioned NEID spectrograph, we constrain the stellar obliquity of \toi via the Rossiter-McLaughlin (RM) effect from both radial velocity (RV) and Doppler Tomography (DT) signals. The 3$\sigma$ upper bounds of the projected stellar obliquity $|\lambda|$ from both models are below 60$^\circ$.
The large host star separation ($a/R_\star {\sim} 17$), combined with the system's young age, makes it unlikely that the planet has realigned its host star. The stellar obliquity measurement of \toi probes the architecture of a young gas giant beyond the reach of tidal realignment ($a/R_\star {\gtrsim} 10$) and reveals an aligned or slightly misaligned system.
\end{abstract}


\section{Introduction} \label{sec:intro}
Stellar obliquity describes the angle between a planet's orbital axis and its host star's spin axis. Giant planets orbiting close to their host stars present a surprisingly wide range of stellar obliquity from zero to 180$^\circ$ \citep[e.g.,][]{albr12}. It is still unclear whether the stellar obliquities reflect close-in giant planets' origin channels -- aligned for in-situ formation or disk migration versus misaligned for high-eccentricity tidal migration (see Section 3.2 of \citealt{daws18} for a review) -- or whether other mechanisms are more important in sculpting the obliquity distribution.
Proposed physical processes include the planet's primordial misalignment of the protoplanetary disk \citep[e.g.,][]{baty12}, the star's magnetospheric interactions with the protoplanetary disk \citep[e.g.,][]{lai11}, and angular momentum transport to the stellar surface by stellar internal gravity waves \citep[e.g.,][]{roge12, roge13}. Moreover, close-in giant planets originating from coplanar high-eccentricity tidal migration \citep{petr15} may be aligned. In addition to these proposed mechanisms, planet-star tidal interactions \changes{may} have altered the obliquity distribution for Hot Jupiter hosts \citep[e.g.,][]{winn10}. Consequently, measuring the obliquities of Warm Jupiters -- orbiting too far from their star to cause tidal realignment ($a/R_\star {\gtrsim} 10$) -- could be essential to disentangle these proposed mechanisms.

The \textit{Transiting Exoplanet Survey Satellite} \citep[\tess;][]{rick15} discovered a large sample of Warm Jupiters around bright stars that are feasible for stellar obliquity measurements using the Rossiter-McLaughlin (RM) effect \citep{ross24, mcla24}. As a planet transits across its host star, it modifies the shape of spectral lines of the star that can be used to infer its positions on the stellar disk relative to the stellar spin axis, and constrain the projected stellar obliquity.
Here we use the newly commissioned \neid spectrograph \citep{NEID_optical} on the 3.5-meter WIYN telescope to conduct the RM-effect measurement of \toi (\tic, $V_{\rm mag} {\sim} 10.9$), the host of a 8.2-day, Saturn-sized planet. The large host star separation (i.e., large $a/R_\star$) of \toib, combined the system's young age, makes it unlikely that the planet has realigned its host star. The stellar obliquity measurement of \toi probes the architecture of a young, warm giant system beyond the reach of tidal realignment.

In Section~\ref{sec:obs}, we present the photometric, high-resolution imaging, and spectroscopic observations of \toi using \tess, \kelt, \alopeke, \pharo, \tres, and \neid. In Section~\ref{sec:stellar}, we model the stellar parameters and estimate the system's age using the stellar rotation period and lithium abundance. In Section~\ref{sec:planet}, we model the planetary parameters from the \tess and ground-based transit light curves (Section~\ref{subsec:transit}) and measure the stellar obliquity of \toi using the RM effect and Doppler Tomography (Section~\ref{subsec:obliquity}). Lastly, in Section~\ref{sec:discussion}, we discuss the implication of the stellar obliquity of \toi and place the target in the context of exoplanetary systems. 

\section{Observations} \label{sec:obs}
\subsection{TESS Photometry} \label{subsec:tess}
The \tess data for \toi are available as 10$\times$10 subimages with 2-minute time sampling, and as part of Full-Frame Images with 30-minute sampling. We obtained 3 sectors of \tess Primary Mission data from 2019-Aug-15 to 2019-Sep-11 (Sector 15) and from 2020-Jan-21 to 2020-Mar-18 (Sectors 21 and 22), and 1 sector of \tess extended mission data from 2021-Jul-23 to 2021-Aug-20 (Sector 41). The target will have at least two more sectors of \tess observations in Sector 48 \changes{(2022-Jan-28 to 2022-Feb-26)} and Sectors 49 \changes{(2022-Feb-26 to 2022-Mar-26)}.

The transit signal was detected with a period of ${\sim} 8.16$ days at high significance independently by the NASA Science Processing Operations Center (SPOC) pipeline \citep{jenk16} and the MIT Quick-Look Pipeline \citep[QLP;][]{huan20b, huan20c}, and was released to the public for follow-up observations as TOI-1268.01. In total, 14 transits of \toib were observed by \tess. The \tess light curves do not show any strong instrument systematics. We used the Pre-search Data Conditioning SAP flux \citep[$\mathtt{PDC\_SAPFLUX}$;][]{Stumpe2012, Stumpe2014} for the light curve analysis.

\begin{figure*}
    \hspace*{0.3cm}
    \includegraphics{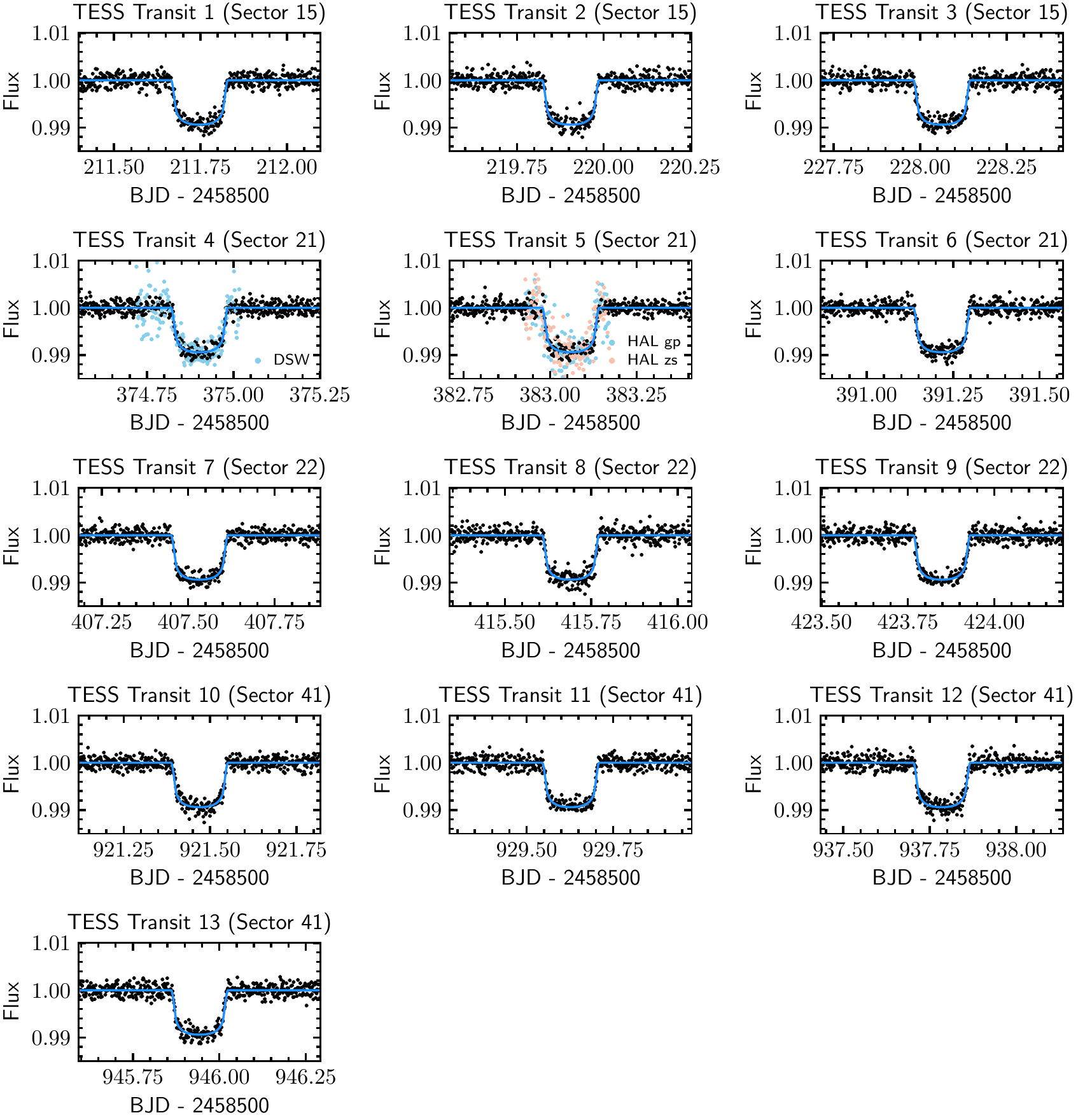}
    
    \vspace*{\dimexpr-\parskip-3.cm\relax}
    \parshape 6
    .35\textwidth .65\textwidth
    .35\textwidth .65\textwidth
    .35\textwidth .65\textwidth
    .35\textwidth .65\textwidth
    .35\textwidth .65\textwidth
    .35\textwidth .65\textwidth
    \makeatletter
    \refstepcounter\@captype
    \addcontentsline{\csname ext@\@captype\endcsname}{\@captype}
    {\protect\numberline{\csname the\@captype\endcsname}{ToC entry}}
    \csname fnum@\@captype\endcsname:
    \makeatother
    Detrended \tess and ground-based transit light curves of \toib. In total we obtained 13 \tess transits in 2-minute cadence from Sectors 15, 21, 22, and 41, and 8 more full or partial transits from ground-based observatories. The Deep Sky West (DSW, simultaneously with \tess Transit 4) and Haleakala (simultaneously with \tess Transit 5) observations are jointly fitted with the \tess transits. The blue curves presents the best-fit transit models.
    \label{fig:lcs}
    \vspace*{1cm}
\end{figure*}

\subsection{Ground-based Transit Photometry} \label{subsec:gb_transits}
Through the TESS Follow-up Observing Program (TFOP) collaboration \citep{collins:2018}, we observed eight full or partial transits of \toib with ground-based seeing-limited telescopes, including three transits observed simultaneously with the \tess observations. We used the $\mathtt{TESS Transit Finder}$, which is a customized version of the $\mathtt{Tapir}$ software package \citep{Jensen:2013}, to schedule our transit observations. These observations confirm that the transiting signal originated from within less than 6\arcsec\ of the target star. Gaia EDR3 \citep{lind21} reports no additional stars within 10\arcsec\ of \toi.

We include two good-quality full transits that were taken simultaneously with \tess in our transit modeling (see Figure \ref{fig:lcs}). The first transit was taken on \ut2020-01-26 by the Deep Sky West 0.5-m telescope near Rowe, NM, USA in the $g'$ band and detected an on-time transit in a 10.9\arcsec\ aperture. The second transit was taken on \ut2020-02-03 simultaneously in $g_p$ and $z_s$ filters with a 6.3\arcsec\ apertures from the Las Cumbres Observatory Global Telescope \citep[LCOGT;][]{Brown:2013} 0.4\,m network node at Haleakala observatory. The light curves were reduced with $\mathtt{AstroImageJ}$ \citep{Collins:2017}.

In addition, we obtained six transits of \toib with full or partial transit baselines from various ground-based facilities. These transit observations played an important role in confirming the transit to be on target and ruling out nearby eclipsing binaries. We do not include these observations in our transit model because of their partial transit baselines or additional complications due to meridian flip (i.e., telescope pointing crossing the meridian during the observation) that introduces systematic flux offset. We list the observations below and these data can be found on ExoFOP website (DOI: \dataset[10.26134/ExoFOP3]{https://doi.org/10.26134/exofop3}).

\begin{itemize}
\item On \ut2020-01-09, an ingress was observed to be on target from the Kotizarovci Observatory 0.3\,m Telescope, near Viskovo, Croatia, in a Baader $R$ 610\,nm longpass filter using a 10.8\arcsec\ aperture, marginally contaminated by a nearby star.

\item On \ut2020-01-10, an egress was observed from the LCOGT 0.4\,m telescope from Teide Observatory in the $z_s$ filter using an uncontaminated 10.2\arcsec\ aperture. 

\item On \ut2020-03-06, the TRAPPIST-North team observed an on-time, almost full transit in the $B$ band. The detection is complicated by a meridian flip at ingress and a strong increase in sky background as the nominal time of egress approached.

\item On \ut2021-04-18, a partial transit was observed in the $B$ band from the OAUV-TURIA1 (0.143\,m) telescope near Valencia, Spain. The detection is complicated by a meridian flip during predicted ingress and strong residuals.

\item Also on \ut2021-04-18, the same transit was observed in a 6.7\arcsec\ uncontaminated aperture in the $B$ band from the Observatory de Ca l'Ou 0.4\,m telescope near Barcelona, Spain. 

\item On \ut2021-04-27, a partial was observed in the $g$ band using an uncontaminated 4.7\arcsec\ aperture from the Wellesley College Whitin Observatory CDK700 telescope near Wellesley, MA. 
\end{itemize}

\subsection{Long-term Photometric Observation} \label{subsec:KELT}
\begin{figure*}
    \centering
    \hspace*{-0.5cm}
    \includegraphics{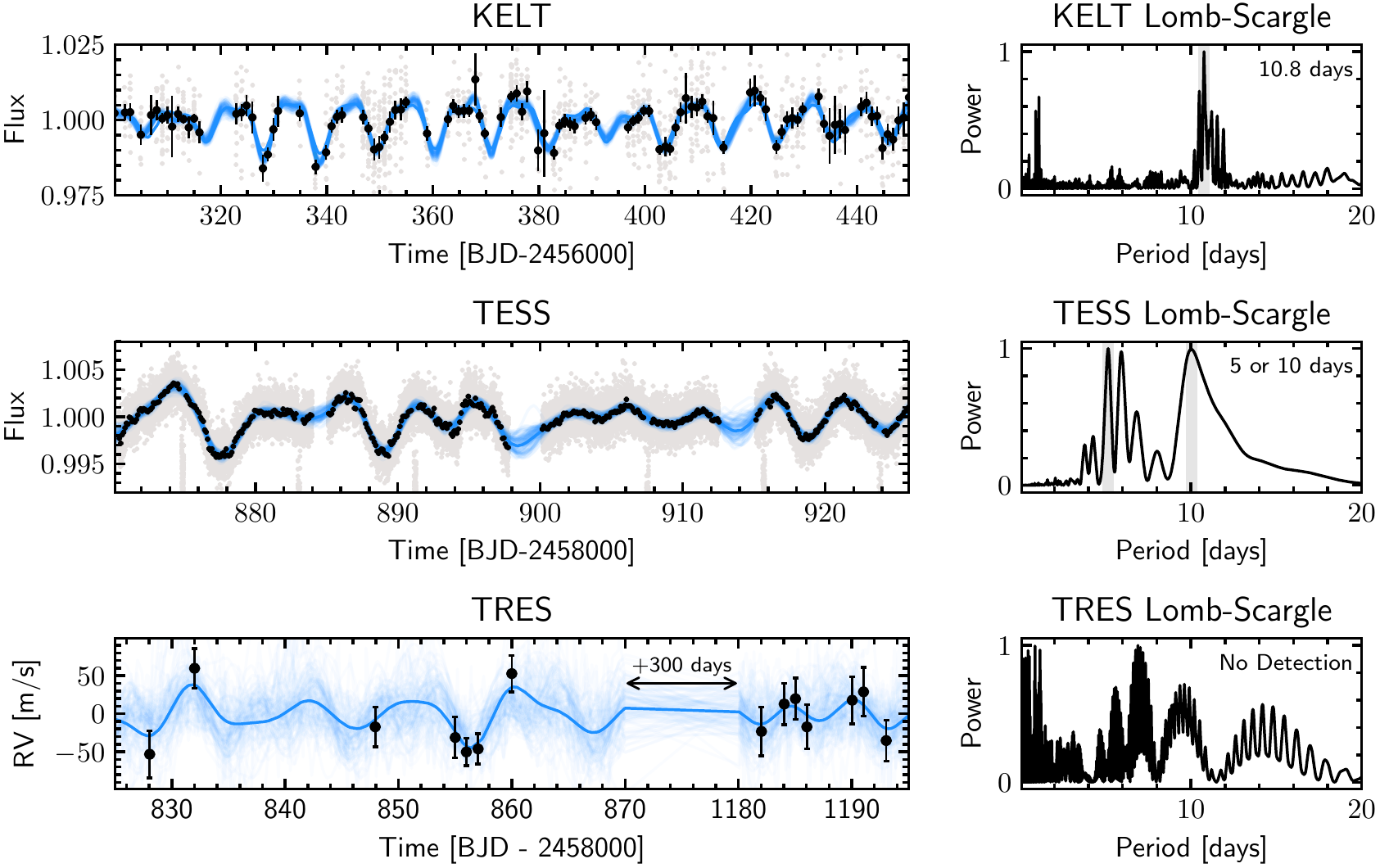}
    \caption{Long-term photometric and spectroscopic observations of \toi using the \kelt, \tess, and \tres. Only a fraction of the \kelt and \tess light curves are presented due to the limited space. The stellar rotation period is clearly detected in \kelt and \tess photometry in the Lomb-Scargle periodogram. Neither stellar rotation period nor planetary orbital period is detected in \tres data. Applying a rotational kernel in Gaussian Process (GP) on \kelt, \tess, and \tres data (see Section~\ref{sec:stellar}), we infer the stellar rotation period as $10.84\pm0.07$ days, consistent with the periodogram results. The blue curves are predicted GP models and light blue curves are drawn from posteriors.}
    \label{fig:stellar}
\end{figure*}

The Kilodegree Extremely Little Telescope (KELT) Survey \citep{pepp03, pepp07} also monitored the star for over two years from BJD 2455976 to BJD 2457022 as part of its normal survey. The precision of the KELT photometry is not sufficient to detect the transit signals. However, the long term monitoring from KELT was used to measure the stellar rotation period. The Lomb-Scargle periodogram \citep{lomb76, scar82, vand18} of the KELT light curve reveals that the star has a rotation period of 10.8 days (Figure \ref{fig:stellar}). This detection helps to break the degeneracy between the rotation period determined from the TESS light curves, which shows two peaks at ${\sim} 5$ days and ${\sim} 10$ days.

\subsection{High-resolution Imaging Observation} \label{subsec:hires}
\begin{figure}
    \hspace*{-0.1cm}
    \includegraphics[width=0.45\textwidth]{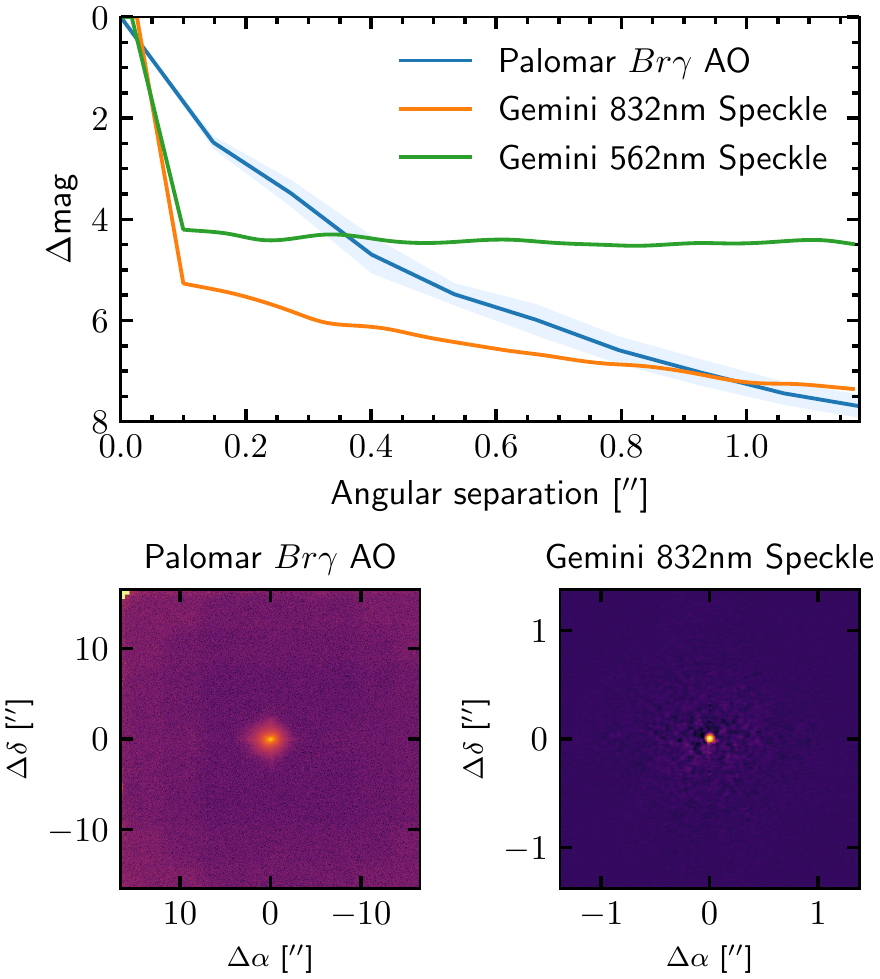}
    \caption{Upper: The 5$\sigma$ contrast curves of \toi from Palomar in $Br\gamma$ filter from adaptive optics (AO) imaging, and from Gemini North in 832nm and 562nm bands from speckle imaging. Lower: The AO image (left) and reconstructed speckle image (right) of \toi. No nearby companions are identified.}
    \label{fig:imaging}
\end{figure}

High-resolution imaging is required to detect nearby companions or background objects that cannot be resolved by seeing-limited photometry. We obtained both adaptive optics (AO) and speckle imaging of \toi, as shown in Figure~\ref{fig:imaging}.
On \ut2020-01-08, the \pharo instrument \citep{hayw01} on Palomar-5m collected AO images of \toi in the narrow-band $Br\gamma$ filter. No companions are identified down to a contrast of 5.481 magnitudes at $0.5\arcsec$.
On \ut2021-02-02, the \alopeke speckle instrument \citep{scot19} on Gemini North-8m took simultaneous speckle imaging in 832\,nm and 562\,nm bands. No companions are detected down to a contrast of 6.36 mag at $0.5\arcsec$.

Although not shown in Figure~\ref{fig:imaging}, we obtained the following observations on the Sternberg Astronomical Institute (SAI)-2.5m telescope located at Mt Shatdzhatmaz in the North Caucasus and on the Shane-3m at Lick Observatory in Mount Hamilton, California, USA. 
On \ut2020-11-29, the Speckle Polarimeter on SAI-2.5m obtained speckle imaging of \toi in $I$ filter.
On \ut2019-11-12, the ShARCS instrument \citep{kupke_et_al2012, gavel_et_al2014} on Shane-3m collected AO images of \toi in $Ks$ and $J$ filters. The ShARCS data were reduced and analyzed using the open-source Python-based SImMER pipeline available on GitHub and described in previous publications \citep{hirsch_et_al2019, savel_et_al2020}.
\toi appeared single in both observations. 

\subsection{Long-term Spectroscopic Observation} \label{subsec:TRES}
We obtained 14 spectra with the Tillinghast Reflector Echelle Spectrograph (\tres) on the 1.5\,m telescope at the Fred Lawrence Whipple Observatory, from UT 2019-12-10 to UT 2020-12-27. \tres has a resolving power of $R\approx44,000$, and covers a wavelength range from 385\,nm to 906\,nm. The spectra were extracted following \citet{2010ApJ...720.1118B}, and radial velocities were measured using a cross correlation analysis against a template spectrum generated from a median combination of all \tres observed spectra \citep{quin12}. We also make use of the \tres spectra to measure the atmospheric parameters of the host star via the Stellar Classifications Pipeline \citep[SPC;][]{buch12, buch14}, finding an effective temperature of $T_\mathrm{eff} = 5288 \pm 50$ K, surface gravity $\log{g}$ of $4.62 \pm 0.10$, and bulk metallicity [M/H] of $+0.16 \pm 0.08$. The projected broadening width $v_{\rm broadening} = 4.1 \pm 0.5$ km\,s$^{-1}$.
The $v_{\rm broadening}$ here does not correct for macroturbulence, so the stellar rotational velocity $v\sin i_\star$ must be smaller than the reported value.

\toi exhibits significant photometric variability due to its youth, and as such we also expect significant jitter in the radial velocities (RVs). The \tres RVs exhibit scatter at the 50 m\,s$^{-1}$ level with a typical RV precision at ${\sim} 30$ m\,s$^{-1}$.
The Lomb-Scargle periodigram of the \tres RVs detects neither the stellar rotation period nor the planetary orbital period due to sparse observations and entangled stellar activity and planetary signals.

\subsection{Transit Spectroscopic Observation} \label{subsec:NEID}
We observed one transit of \toib with the extremely high precision NEID spectrograph \citep{NEID_optical, NEID_budget} on the 3.5\,m WIYN telescope at the Kitt Peak National Observatory (KPNO) in Arizona, USA.
NEID is a fiber-fed \citep{kano18}, actively environmentally stabilized spectrograph \citep{robe19, stef16} with a resolution of $R\approx110,000$ and a wavelength coverage of 380\,nm to 930\,nm. The observation was taken on UT 2021-05-04 during the transit of \toib, and covered about 1.5-hour baseline before the transit. We used an exposure time of 8 minutes for each observation, and in total obtained 37 spectra. The spectra were extracted and radial velocities were reduced by the \neid standard data reduction pipeline $\mathtt{NEID}$-$\mathtt{DRP\,v1.1.2}$\footnote{https://neid.ipac.caltech.edu/docs/NEID-DRP}, which derives cross correlation based RVs (we used $\mathtt{CCFRVMOD}$ data produced by the pipeline), and separately by the $\mathtt{SERVAL}$ pipeline \citep{zech18}, which derives RVs using the reconstructed stellar template from observations (see Section 3.1 in \citealt{stef21} for the \neid customization). 
The two pipelines derive similar RVs that are consistent within 1$\sigma$ uncertainties except a few data points. The achieved median, \changes{photon-limited} RV precision for both pipelines is ${\sim} 5.8$ m\,s$^{-1}$. Reduced RVs are presented in Figure~\ref{fig:rm}.

To directly measure the Doppler shadow cast by the planet on the spectroscopic line profiles of the star, we perform a least-squares deconvolution \citep{donati97} between the \neid spectra and a synthetic non-rotating spectral template. The synthetic template is generated using a set of ATLAS-9 atmosphere models \citep{Castelli:2004} at the stellar atmosphere parameters of \toi. The line profiles are computed for each order of an observation, and weighted-average combined into a single line profile per epoch. Section~\ref{subsec:obliquity} describes the modeling of the line profiles to retrieve the planetary orbital obliquity, and Figure~\ref{fig:rm} shows the tomographic shadow of the planetary transit. 

\section{Stellar Properties} \label{sec:stellar}
\subsection{SED Modeling}
We use $\mathtt{astroARIADNE}$\footnote{https://github.com/jvines/astroARIADNE} to model the spectra energy distribution (SED) of the star. We use the three Gaia band magnitudes, three 2MASS band magnitudes, and the four WISE band magnitudes in the modeling, and use the Gaia parallax, $T_{\rm eff}$ and [M/H] derived from the \tres spectra as our priors. The uncertainties of the photometry bands are inflated following methods described in $\mathtt{EXOFASTv2}$ \citep{Eastman:2013, Eastman:2019}. We use the PHOENIX models and MIST isochrones in the SED modeling. The best fitted stellar parameters and their uncertainties are $R_\star=0.86\pm0.02\,\rsun$, $M_\star=0.9\pm0.13\,\msun$, $T_{\rm eff} = 5257\pm40$ K, $\log{g} = 4.52\pm0.07$, and [M/H]$= +0.17\pm0.06$.  

\subsection{Stellar Rotation}
As discussed in Section~\ref{subsec:tess} and \ref{subsec:KELT}, the star exhibits a clear rotation signature in both the \tess and \kelt photometry. We use a Gaussian Process (GP) model with a rotation kernel to infer the rotation period of \toi. The rotation kernel is composed of two damped harmonic oscillators with the characteristic frequencies of $1/P$ and $2/P$ to model stellar variability at the rotation period itself and at harmonics. Five free parameters in the rotation term are the rotation period $P$, two quality factors $Q_0$ and $dQ$ describing the damping timescales of each oscillator, and $\sigma$ and $f$ describing the amplitudes of each oscillator.
We apply the kernel to the \kelt, \tess, and \tres observations using the $\mathtt{celerite2}$ package \citep{celerite1, celerite2}. Transits are masked from \tess light curves.
We run MCMC to sample posteriors using the $\mathtt{PyMC}$ package \citep{exoplanet:pymc3}. We sample four chains, each with 10,000 burn-in steps, 3,000 draws, and use a target acceptance rate of 0.95. We assess the MCMC convergence using the Gelman-Rubin diagnostic ($\hat{R} < 1.01$ for convergence) and find all the inferred parameters have $\hat{R} \le 1.001$.  
In Figure~\ref{fig:stellar}, we present the flux and RV variations predicted by GP models in blue curves and draws from the posteriors in light blue curves. The GP models perform well on predicting flux variations on \tres and \tess light curves, whereas perform poorly on \tres RVs due to the sparse sampling on \tres data and complication from planetary signal. We test the GP models with and without \tres RVs and find similar rotation period posteriors.
The inferred rotation period for \toi is $P_{\rm rot} = 10.84 \pm 0.07$ days. Combining the stellar rotation period and radius, the equatorial rotational velocity of \toi is $v_{\rm rot} = 2\pi R_\star/P_{\rm rot} = 4.02 \pm 0.10$ km\,s$^{-1}$.

The inclination of the star $i_\star$ can be inferred from the projected rotational velocity of the host star and its equatorial rotational velocity \citep{masu20}. Since we only know the projected broadening width ($v_{\rm broadening} = 4.1 \pm 0.5$ km\,s$^{-1}$), the true projected rotational velocity could be smaller and this difference could lead to an overestimation of the inclination. We apply priors on $P_{\rm rot}$ and $ R_\star$ from stellar fits, as well as a uniform prior on $\cos{i_\star}$, and infer $i_\star = 76 \pm 10 ^\circ$ using the MCMC. 

\subsection{Stellar Age}

\begin{figure*}
\gridline{\fig{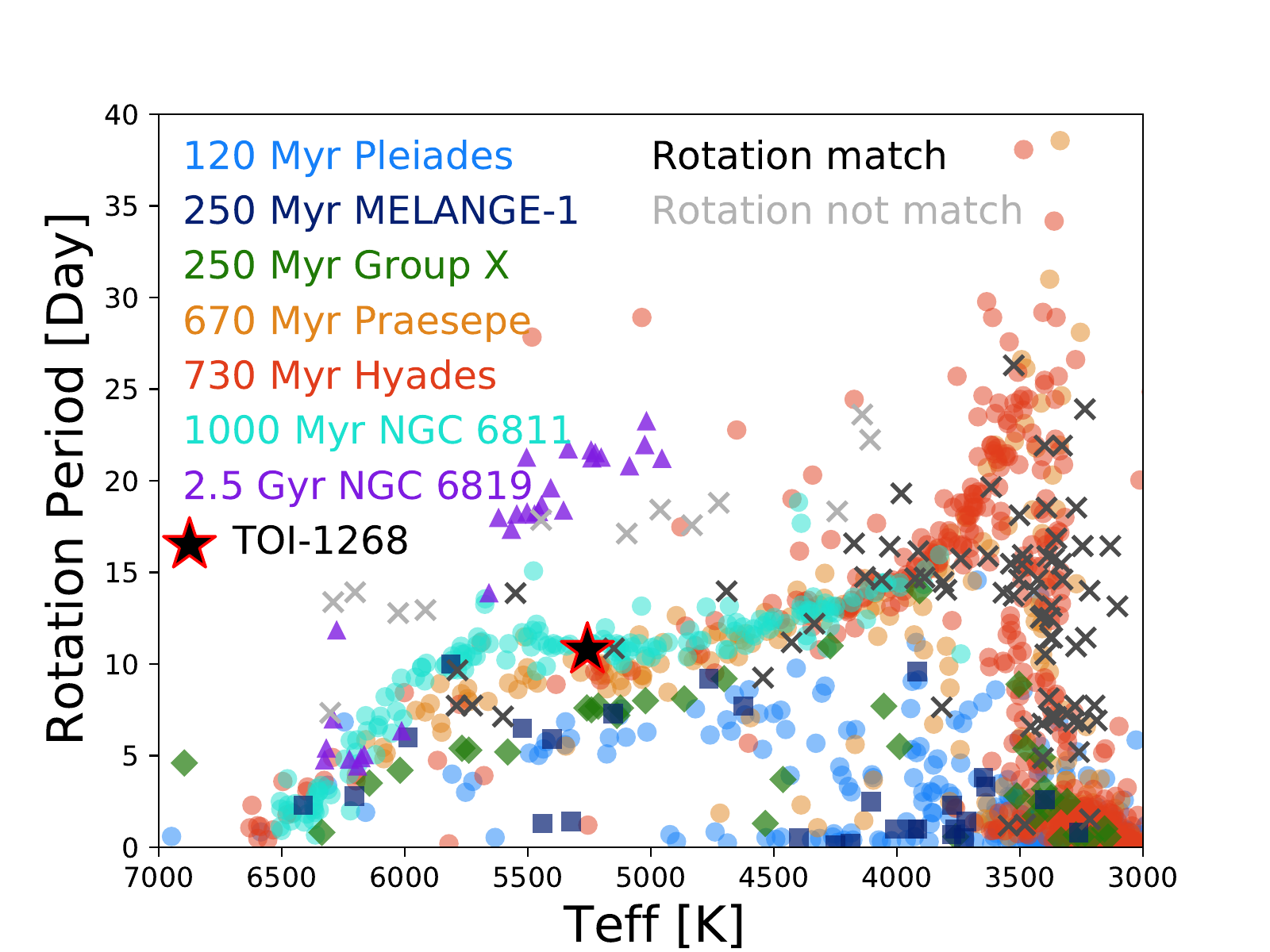}{0.48\textwidth}{(a) Rotation period vs. effective temperature}
          \fig{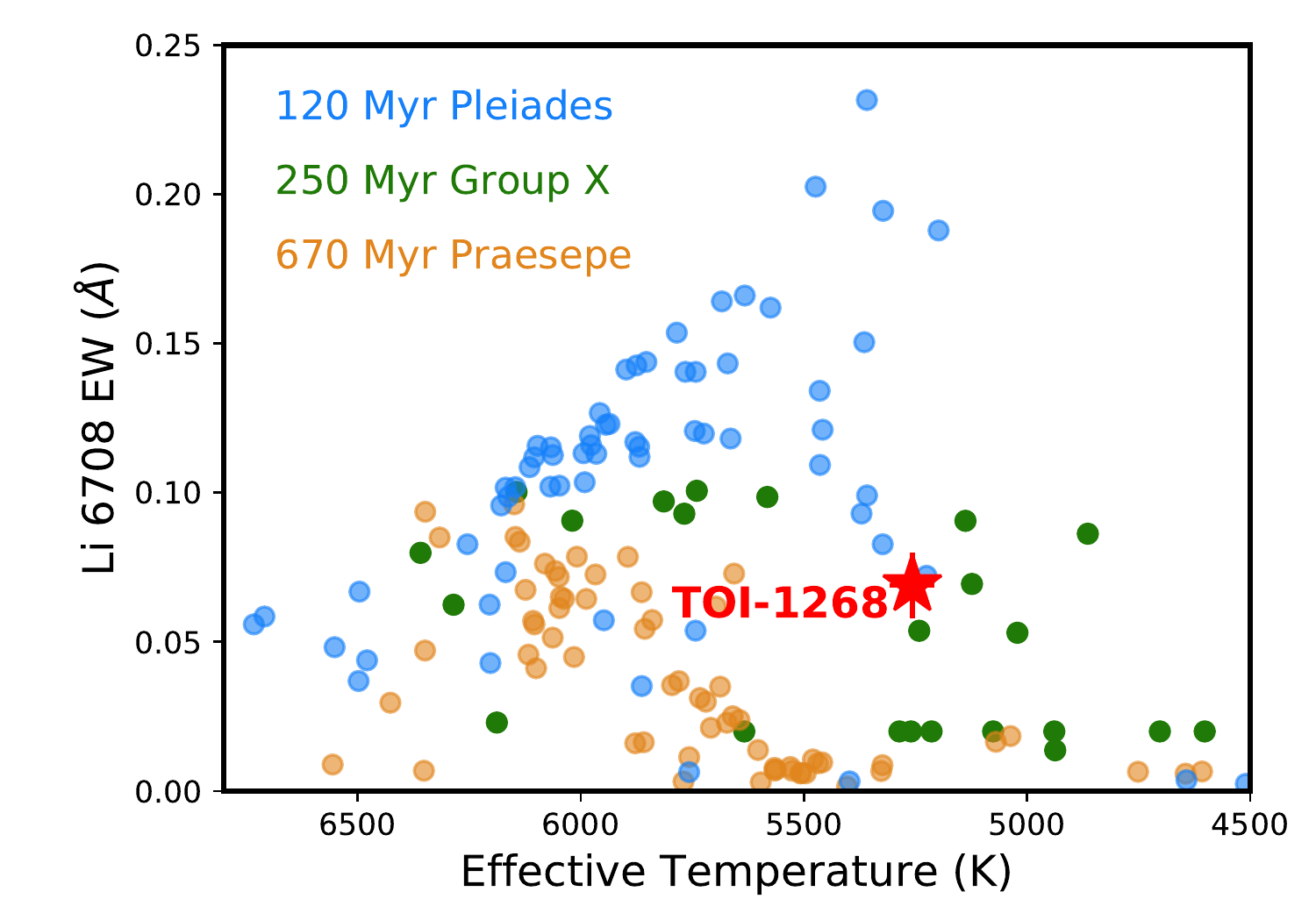}{0.48\textwidth}{(b) Lithium Equivalent Width (EW) vs. effective temperature}}
\caption{(a) Rotation period vs. effective temperature of \toi (black star with red edge) and its neighboring stars in comparison with clusters groups with well determined ages. The clusters we used are 120 Myr Pleiades (blue circles, \citealt{Rebull:2016}), 250 Myr MELANGE-1 (blue squares, \citealt{toff21}), 250 Myr Group X (green diamonds, Newton et al. in prep), 670 Myr Praesepe and 730 Myr Hyades (orange and red circles, \citealt{Douglas:2016, Douglas:2019}), 1 Gyr NGC 6811 (cyan circles, \citealt{Curtis:2019}), and 2.5 Gyr NGC 6819 (purple triangles, \citealt{Meibom:2015}). For stars we identified with the $\mathtt{Comove}$ package, the ones with rotation period and effective temperature consistent with \toi are plotted with dark grey crosses, while the ones deviate from it are plotted with light grey crosses. (b) Lithium equivalent width (EW) vs. effective temperature of \toi (red star) and its neighboring stars in comparison with clusters groups with well determined ages. The clusters we used are 120 Myr Pleiades (blue circles), 250 Myr Group X (green circles, Newton et al. in prep), 670 Myr Praesepe (orange circles).
 \label{fig:age}}
\end{figure*}

\toi does not belong to any known association based on a search of the BANYAN $\Sigma$ catalog \citep{Gagne:2018}. We used the $\mathtt{Comove}$ package\footnote{https://github.com/adamkraus/Comove} \citep{toff21} to query Gaia EDR3 \citep{lind21} and search for associations within 50\,pc, and did not find any clear clustering in velocity space. We identified 95 candidate stars that are brighter than \tess magnitude of 13.5, and could be associated with \toi, and use the Discrete Fourier Transform (DFT) clean algorithm in $\mathtt{VARTOOLS}$ \citep{hart16} to measure their rotation periods with \tess FFI light curves. The relation between the rotation period and effective temperature in comparison to clusters with well determined age are shown in Figure \ref{fig:age} (a). The effective temperatures were obtained from the TIC-v8 catalog \citep{stas19}. While some of the candidates have rotation periods consistent with being young (dark grey crosses), many do not (light grey crosses). A full vetting of candidates would be required to use them to further refine the age of \toi.
From the rotation period only, \toi's age is most likely to be between the 120\,Myr Pleiades (colored in blue), and the 2.5\,Gyr NGC 6819 (colored in purple). 

For K-type stars like \toi, lithium is expected to be depleted when the star is older than Praesepe/Hyades ages \citep[e.g.,][]{boes16, cumm17}. However, Li I 6707.8\,nm is clearly detected in both the \tres and \neid spectra. The equivalent width (EW) is measured to be $0.069\pm0.011$\,\AA\,from the \tres data. In Figure~\ref{fig:stellar} (b), we compare the Li abundance of \toi to Pleiades (120 Myr), \changes{Group X (250 Myr; Netwon et al. in prep)}, and Praesepe (670 Myr) clusters. The Li measurements for Pleiades and Praseape are obtained from \cite{Zhou:2021}, where the spectra were obtained as part of the long term radial velocity surveys on the TRES spectrograph by \cite{quin12} and \cite{quin14}. The Li abundance of \toi is richer than Praesepe and in agreement with Pleiades. Combining \changes{Li and stellar rotation period} information, \toi's age is likely between Pleiades and Praesepe clusters, i.e., 120--670 Myr.

We use $\mathtt{BAFFLES}$ \citep{stan20}, a package that uses empirically determined relations to compute age posteriors for field stars from measurements of $\log R^{\prime}_{\rm HK}$ Calcium emission or lithium equivalent width absorption and B-V color to estimate the age of \toi. \changes{From the TRES spectra, we measure $\log R^{\prime}_{\rm HK}=-4.3\pm 0.19$.} Since the B-V color from the catalog has relatively large error bars, we use MIST isochrones and the best fitted SED of \toi to derive a more accurate B-V color of $0.83\pm0.03$. The ages independently estimated from the Calcium emission and Lithium lines are consistent with each other. The Calcium age posterior gives 130 Myr--1.4 Gyr in the 1$\sigma$ credible interval (CI). The lithium age posterior gives 220--500 Myr in the 1$\sigma$ CI. The combined posterior estimates the age of \toi is 190 Myr--370 Myr in the 1$\sigma$ CI (or 76 Myr--600 Myr in the 2$\sigma$ CI). 

Using the above information, we conclude that the rotation, lithium abundance, and activity index all give consistent ages, and confirm the youth of \toi. 

\section{Planetary Properties} \label{sec:planet}
\subsection{Transit Model} \label{subsec:transit}
We use a quadratic limb darkening transit model \citep{mand02, exoplanet:kipping13} plus a rotational Gaussian Process kernel \citep{celerite1, celerite2} to model the transit light curves and the rotational modulation introduced by stellar activity. We perform the light curve fit using the \tess 2-minute cadence data only and also the \tess data jointly with two ground-based transits described in Section~\ref{subsec:gb_transits}. To reduce the computational time, we trim the \tess light curves to roughly three times the transit duration before the ingress and after the egress. No transit-timing variations on \toib are detected in a preliminary light curve fit. Because of that, we directly model the orbital period $P$ and the reference mid-transit time $T_C$. Free parameters in our model include $\big\{\rho_{\textrm{circ}}, b, r_{\textrm{p}}/r_{\star}, P, T_C\big\}$, the quadratic limb darkening parameter $\big\{u_0, u_1\big\}$, and GP parameters for the rotational kernel (see Section~{\ref{sec:stellar}}). We take the GP parameters derived from the out-of-transit \tess data as priors.
Here we model $\rho_{\textrm{circ}}$, the stellar density of the host star assuming zero eccentricity, and later compare it to the $\rho_\star$ from isochrone fitting to infer the planet's eccentricity $e$ and argument of periapse $\omega$. To jointly model the \tess and ground-based light curves, we use an independent pair of limb-darkening parameters for each filter and separate GP models for \tess and ground-based transits due to different cadences.

In Figure~\ref{fig:lcs}, we present the detrended \tess and ground-based transit light curves from a joint fit. The orbital period and transit ephemeris of the planet are tightly constrained. A summary of planetary parameters can be found in Table~\ref{tbl:parameters}.

\begin{figure*}
\gridline{\fig{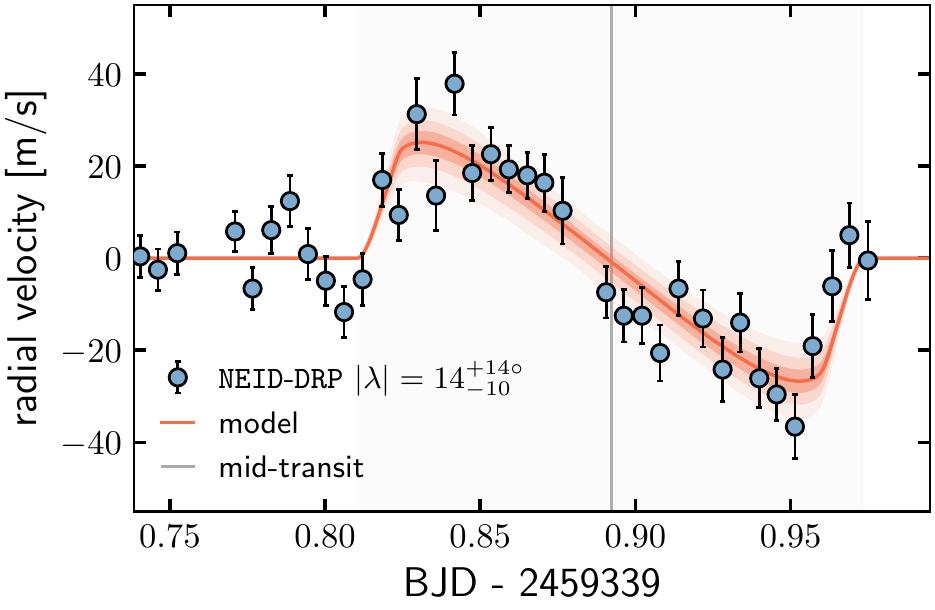}{0.45\textwidth}{\hspace*{1cm}(a) Radial velocities reduced by the $\mathtt{NEID}$-$\mathtt{DRP\,v1.1.2}$ pipeline}
          \fig{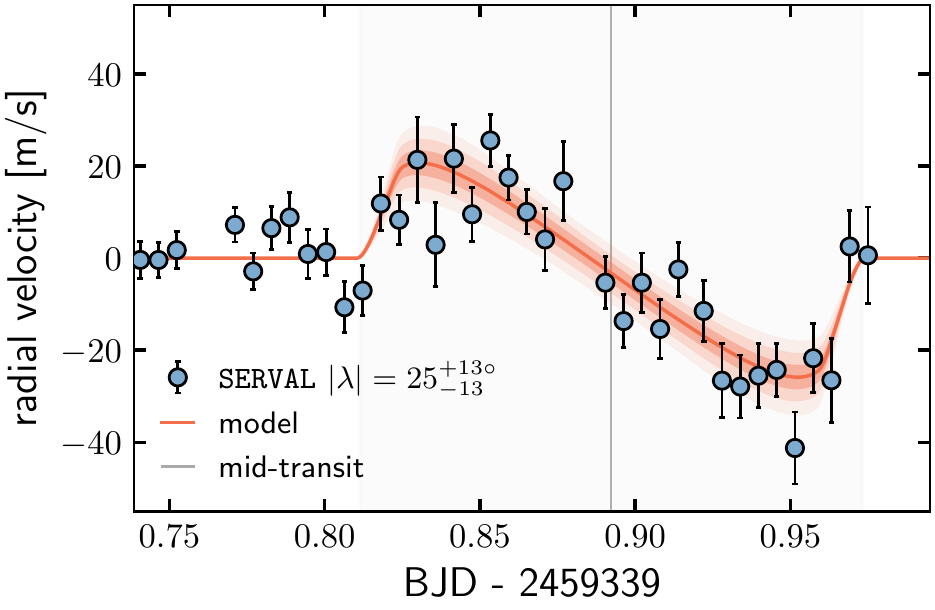}{0.45\textwidth}{\hspace*{1cm}(b) Radial velocities reduced by the $\mathtt{SERVAL}$ pipeline}}
\gridline{\fig{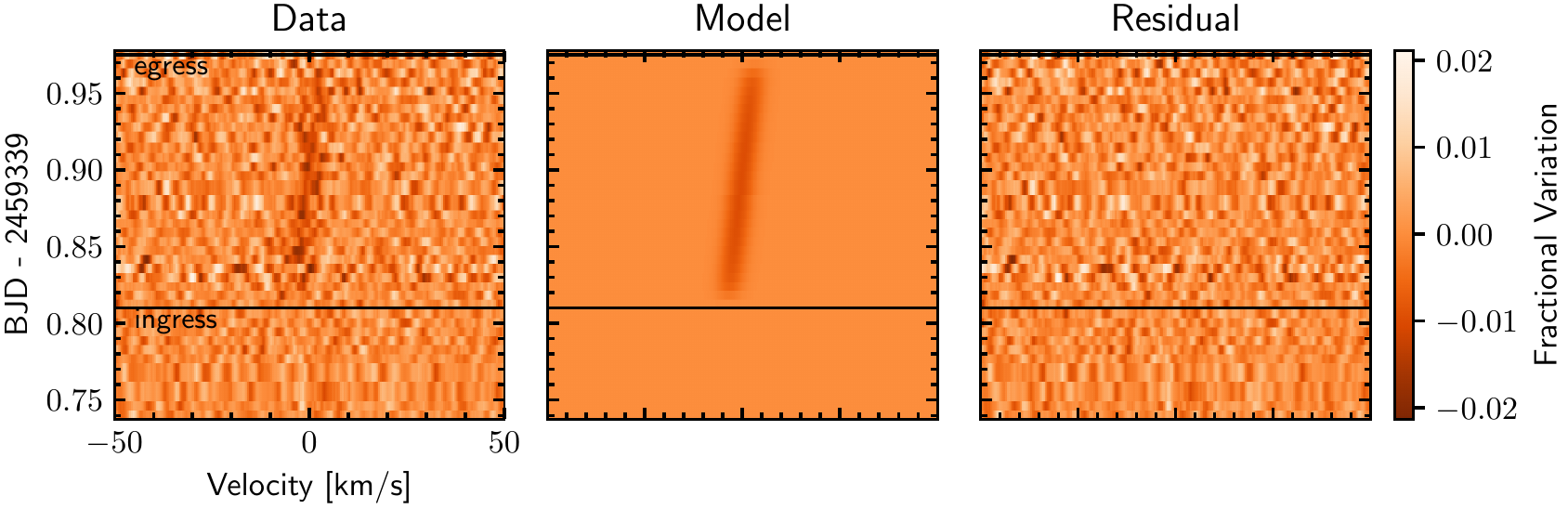}{0.95\textwidth}{(c) Doppler Tomography signals, from which we infer $|\lambda| = 40^{+7\circ}_{-10}$}}
\caption{(a) and (b) In-transit radial-velocity measurements of the \toi system using the \neid spectra. The blue dots and black bars are \neid RVs and their corresponding uncertainties. Using the Rossiter-McLaughlin effect, the projected stellar obliquity is constrained. (c) The Doppler Tomography \changes{signal} of the \toi system during \toib's transit. The left, middle, and right panels are data extracted from the \neid spectra, best-fit model, and the residual of the data after subtracting the best-fit model. The colorscale presents the flux variation of the velocity channel. We expect to observe a decrease in flux in the velocity channel of the star blocked by the planet. \label{fig:rm}}
\end{figure*}

\begin{figure*}[htb!]
    \centering
    \includegraphics{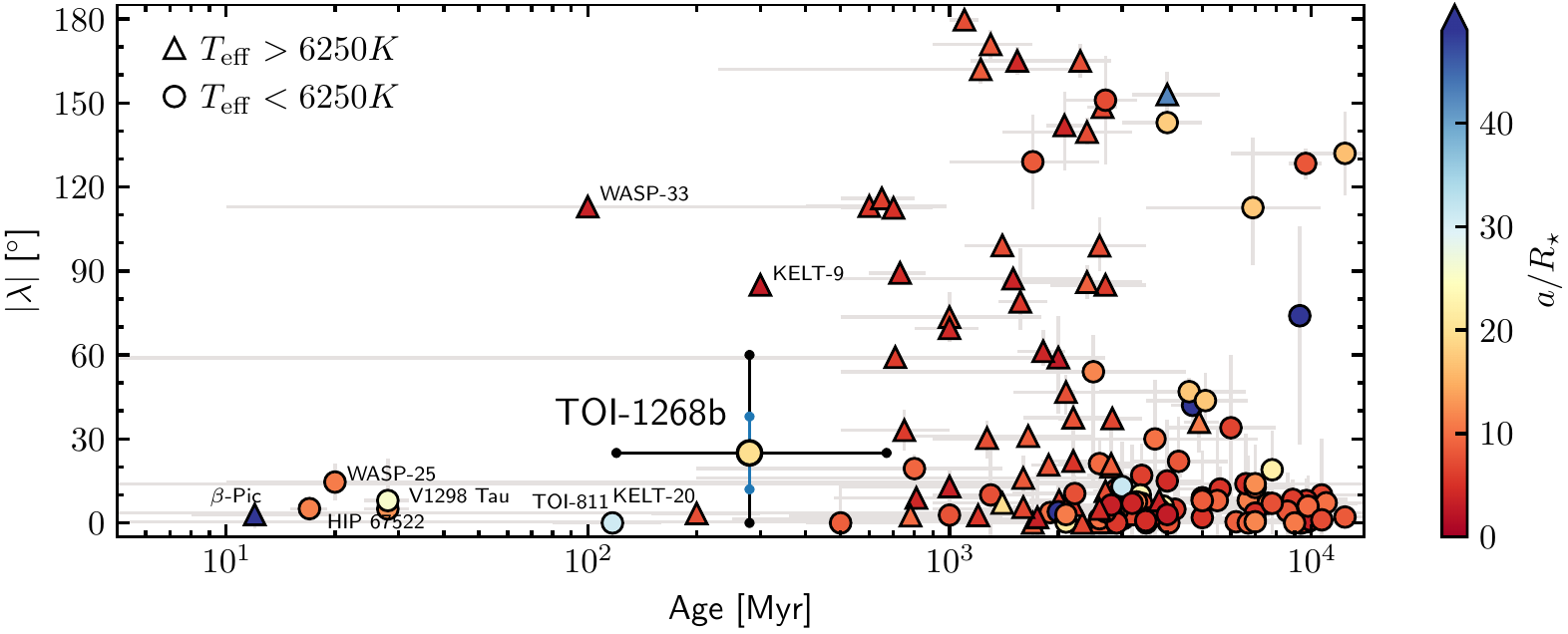}
    \caption{Projected stellar obliquity versus estimated stellar age for all Hot/Warm Jupiters. Planets are colored by $a/R_\star$ and labeled in triangles or circles given their host star effective temperatures (i.e., triangles for host stars above the Kraft break $T_\mathrm{eff} > 6250 K$ and circles for stars below the Kraft break). Grey lines show 1$\sigma$ uncertainties. For \toi, we plot its estimated age between the 120 Myr Pleiades and 670 Myr Praesepe clusters and its stellar obliquity using representative values inferred from $\mathtt{SERVAL}$ RVs labeled in blue error bars. The 3$\sigma$ credible interval of obliquity posteriors derived from the RV and DT signals are shown in black.}
    \label{fig:obl}
\end{figure*}

\subsection{Stellar Obliquity Measurement} \label{subsec:obliquity}
We use the \neid RM-effect signals to infer the stellar obliquity from two separate approaches: (1) model the RV anomalies reduced by the $\mathtt{NEID}$-$\mathtt{DRP\,v1.1.2}$ and the $\mathtt{SERVAL}$ pipelines (see Section~\ref{subsec:NEID} for the description of the pipelines), and (2) directly model the planetary shadow extracted from the spectra using the planet's transit \citep[i.e., Doppler Tomography or DT;][]{coll10}. For both approaches, we jointly model the RM-effect signals with the \tess transit light curves. Doing so allows one to solve the complex covariances between the impact parameter $b$, the projected stellar obliquity $\lambda$, and the projected stellar rotation velocity $v\sin i_\star$.

For the RV fits, our model includes parameters mentioned in the transit model (Section~\ref{subsec:transit}), and also the $\lambda$, $v\sin i_\star$, and a RV jitter term $\sigma_{\rm RV}$ as free parameters. We place uniform priors on these parameters.\footnote{We also tested placing a prior on $v\sin i_\star$ based on the observed line broadening, and found minimal changes on posteriors.}
The RV anomaly due to the transit is modeled using the $\mathtt{starry}$ package \citep{exoplanet:luger19}, which takes the analytical expression of the radial velocity of a stellar disk \citep{shor18} and converts the polynomials to spherical harmonic coefficients. The calculated RVs do not account for macroturbulence or instrumental broadening. We incorporate $\mathtt{starry}$ into $\mathtt{exoplanet}$, build and sample the joint model using the $\mathtt{PyMC}$. We also add a quadratic trend to model the baseline RV trend introduced by either the planet's orbit or stellar activity. In total, we sample four chains, each with 20,000 tuning steps and 5,000 draws. A target acceptance rate of 0.95 was used. All four chains are confirmed to be converged and the inferred parameters have $\hat{R} \le 1.001$. The quadratic coefficients are consistent with zeros.

For the DT fit, similar to the RV fits, we incorporate the DT data into $\mathtt{exoplanet}$, and build and sample the joint model using the $\mathtt{PyMC}$. At each observing time $t$, we calculate the planet's position on the stellar disk, assuming the star rotates as a rigid body, and identify the stellar velocity channels being blocked by the planet, $v_t$. To model the planetary shadow, we use a Gaussian distribution that centers at $v_t$ and has a standard deviation of $\sqrt{v_{\rm res}^2 + v_{\rm macro}^2}$, where $v_{\rm res}$ is the velocity resolution set by the spectrograph resolution and $v_{\rm macro}$ is the macroturbulence velocity determined by the host star. The planetary shadow is further scaled with the photometric flux at time $t$ and normalized by the total stellar velocity flux over the planetary velocity flux. We sum up the likelihoods of the DT signals at all observing times, and infer the planet's orbital orientation, along with its \tess transit light curves. The $\mathtt{PyMC}$ setup and the convergence test are the same as the ones described in the RV fits.

In Figure~\ref{fig:rm} (a) and (b), we present the \neid RVs reduced by the $\mathtt{NEID}$-$\mathtt{DRP\,v1.1.2}$ and $\mathtt{SERVAL}$ pipelines, and their corresponding RM-effect models and uncertainties. In Figure~\ref{fig:rm} (c), we show the DT data (left panel), the best-fit model (middle panel), and the residual of the data after subtracting the model (right).
In all three inference models, the projected stellar obliquity $|\lambda|$ posteriors extend from zero to 60$^\circ$ (3$\sigma$ CIs). A polar or retrograde solution of \toi system can be ruled out.
However, the differences in $|\lambda|$ posteriors from three models are still noticeable. In Table~\ref{tbl:parameters}, we summarize the fitting parameters. The $\mathtt{NEID}$-$\mathtt{DRP}$ RVs suggest an aligned system with $|\lambda| = 14^{+14\circ}_{-10}$, whereas the $\mathtt{SERVAL}$ RVs and DT data suggest a slightly misaligned system (i.e, $|\lambda| = 25^{+13\circ}_{-13}$ for $\mathtt{SERVAL}$ and $|\lambda_{\rm DT}| = 40^{+7\circ}_{-10}$ for DT). A possible explanation for the high stellar obliquity inferred by the DT model is the stellar obliquity and impact parameter degeneracy: low impact parameters ($b {\sim} 0.1$) correlate to high stellar obliquities ($\lambda {\sim} 40^\circ$), and slightly higher impact parameters ($b {\sim} 0.2$) correlate to lower stellar obliquities ($\lambda {\sim} 20^\circ$). Since a low impact parameter solution is suggested by the DT model ($b =$ \DTb), we derive the high stellar obliquity solution. The DT inferred impact parameter is still consistent with the one from transit-only fit ($b = 0.2\pm 0.1$). Breaking the degeneracy between the impact parameter and stellar obliquity will be the key step towards precise stellar obliquity measurement in future observations.

Lastly, we use the inclination of the host star ($i_\star = 76 \pm 10 ^\circ$) and the projected stellar obliquity to estimate the true stellar obliquity $\psi$ of \toi. Using the spherical law of cosines, $\cos{\psi} = \cos{|\lambda|}\sin{i}\sin{i_\star} + \cos{i}\cos{i_\star}$, where $i$ is the orbital inclination, we get $\psi = 22.7 \pm 10.5 ^\circ$ for $\mathtt{NEID}$-$\mathtt{DRP}$ RVs, $\psi = 30.4 \pm 11.1 ^\circ$ for $\mathtt{SERVAL}$ RVs, and $\psi = 40 \pm 10 ^\circ$ for the DT signal.

\section{Results and Discussion} \label{sec:discussion}
\toi is an early K-type dwarf hosting an 8.2-day, Saturn-sized planet. Using the host star's rotation period and lithium abundance, we estimated the age of \toi between the 120 Myr Pleiades and the 670 Myr Praesepe or 730 Myr Hyades.
\toib was discovered during the \tess prime mission, and validated by seeing-limited photometry, reconnaissance spectroscopy on \tres, and high-resolution imaging on \alopeke, \pharo, Speckle Polarimeter on SAI-2.5m, and ShARCS. We confirmed the planet using the newly commissioned \neid spectrograph via the RM-effect. The planetary nature of \toib has also been independently confirmed by the KESPRINT consortium through high-precision RV follow-up observations (\v{S}ubjak et al. submitted). 

Using the \neid spectra, the stellar obliquity of \toi was constrained. The stellar obliquity and impact parameter degeneracy and the small $v\sin{i_\star}$ of \toi make it challenging to measure the stellar obliquity precisely from a single transit observation. However, a stellar obliquity greater than 60$^\circ$ can be ruled out at 3$\sigma$ level. The stellar obliquity of \toi is either aligned, suggested by the $\mathtt{NEID}$-$\mathtt{DRP}$ RVs, or slightly misaligned, suggested by the $\mathtt{SERVAL}$ RVs and the DT signal. Further transit spectroscopy observations of the system will be required to resolve the minor discrepancy between the models and refine the stellar obliquity measurement. \cite{osha18} discussed how star spots could compromise stellar obliquity measurements, which further motivate multiple RM-effect measurements on young \toi.
\toi is one of the few studies constraining the stellar obliquity using multiple techniques \citep[see also][]{knud21}, and one of the first studies modeling DT signals on a spotty young star with a high precision stabilized spectrograph. Previous works have made use of lower precision spectrographs that make such a comparison impossible.

The eccentricity of \toib inferred purely from the transit light curves and the stellar density is consistent with a circular or low eccentricity orbit planet. Given \toib's large orbital distance ($a_p = 0.068 \pm 0.02$ au), it will require high tidal dissipation efficiency and/or a nearby companion still coupled and driving eccentricity oscillations of \toib, if the planet has undergone or is undergoing high-eccentricity tidal migration. \toib is likely an outcome of disk migration or in-situ formation.
The large planet-star separation ($a/R_\star {\sim} 17$), along with the system's young age, makes it unlikely to align with its host star by planet-star tidal interactions. The stellar obliquity of the system probes the primordial spin-orbit angles for Warm Jupiters formed in situ or via disk migration and points to an aligned or slightly misaligned system. Strong primordial misalignment, such as by chaotic accretion \citep{bate10}, magnetic warping \citep{lai11}, or an inclined stellar/planetary companion \citep{baty12}, probably did not occur in the system.

In Figure~\ref{fig:obl}, we present the projected stellar obliquity versus stellar age for all Hot/Warm Jupiters for which obliquity measurements are available (data from Albrecht et al., in prep). Planets are colored by their planet-star separations ($a/R_\star$) and circles (triangles) indicate host star temperatures above (below) to the Kraft break (6250 K). \toib stands out for its young age and large planet-star separation.
Currently, giant planets in systems younger than 100 Myr -- $\beta$-Pic b \citep{hira20}, WASP-25b \citep{brow12}, HIP 67522b \citep{heit21}, and V1298 Tau b and c \citep{john21, fein21} -- are all found in aligned systems. \toib might lie in the transitional region where more misaligned systems get discovered. More stellar obliquity measurements of young systems are encouraged before interpreting the observations theoretically.

\begin{table*}
\centering
\tabletypesize{\small}
\caption{Median values and 68\% credible intervals for the stellar and planetary parameters of the \toi (\tic) system. \label{tbl:parameters}}
\begin{tabular}{llccc}
  \hline
  \hline
Parameter & Units & Values \\
\hline\\\multicolumn{2}{l}{Stellar Parameters}&\smallskip\\
~~~~$M_*$\dotfill & Mass (\msun)\dotfill & \mstar\\
~~~~$R_*$\dotfill & Radius (\rsun)\dotfill & \rstar\\
~~~~$\rho_*$\dotfill & Density (cgs)\dotfill & \rhostar\\
~~~~$\log{g}$\dotfill & Surface gravity (cgs)\dotfill & \logg\\
~~~~$T_{\rm eff}$\dotfill & Effective temperature (K)\dotfill & \teff\\
~~~~$[{\rm M/H}]$\dotfill & Bulk metallicity (dex)\dotfill & \feh\\
~~~~$P_{\rm rot}$\dotfill & Stellar rotation period (day)\dotfill & $10.84 \pm 0.07$\\
~~~~$v_{\rm rot}$\dotfill & Equatorial velocity ($\mathrm{km\,s}^{-1}$)\dotfill & $4.02 \pm 0.10$\\
~~~~$G$\dotfill & Gaia $G$ magnitude (EDR3; \citealt{lind21})\dotfill & $10.690 \pm 0.001$\\
~~~~$B_{\mathrm{P}}$\dotfill & Gaia $B_{\mathrm{P}}$ magnitude (EDR3; \citealt{lind21})\dotfill & $11.131 \pm 0.002$\\
~~~~$R_{\mathrm{P}}$\dotfill & Gaia $R_{\mathrm{P}}$ magnitude (EDR3; \citealt{lind21})\dotfill & $10.089 \pm 0.001$\\
\\
\hline\\\multicolumn{2}{l}{Planetary Parameters (transit$+$RM-effect joint model)}\smallskip\\
~~~~$P$\dotfill & Period (days)\dotfill & \per\\
~~~~$T_C$\dotfill & Mid-transit time (BJD)\dotfill & \midt\\
~~~~$R_p/R_\star$\dotfill & Planet-star radius ratio \dotfill & \rprs\\
~~~~$R_p$\dotfill & Radius (\rj)\dotfill & \rp \smallskip\\
&& With $\mathtt{NEID}$-$\mathtt{DRP}$ RVs & With $\mathtt{SERVAL}$ RVs & With DT\smallskip\\
~~~~$\rho_{\rm circ}$& Stellar density assuming the planet has a circular orbit (cgs)\dotfill & \DRPrhocirc & \SERVALrhocirc & \DTrhocirc\\
~~~~$a/R_\star$\dotfill & Planet-star separation\dotfill & \DRPaor & \SERVALaor & \DTaor\\
~~~~$a$\dotfill & Semi-major axis (au)\dotfill & \DRPap & \SERVALap & \DTap\\
~~~~$b$\dotfill & Transit impact parameter \dotfill & \DRPb & \SERVALb & \DTb\\
~~~~$i$\dotfill & Inclination ($^\circ$)\dotfill & \DRPincl & \SERVALincl & \DTincl\\
~~~~$|\lambda|$\dotfill & Projected stellar obliquity ($^\circ$)\dotfill & \DRPlam & \SERVALlam & \DTlam \\
~~~~$v\sin i_\star$\dotfill & Rotational line broadening ($\mathrm{km\,s}^{-1}$)\dotfill & \DRPvsini & \SERVALvsini & \DTvsini\\
~~~~$e$\dotfill & Eccentricity\dotfill & \DRPecc & \SERVALecc & \DTecc\\
~~~~$\omega$\dotfill & Argument of periapse ($^\circ$)\dotfill & \DRPomg & \SERVALomg & \DTomg\\
~~~~$\sigma_{\rm RV}$\dotfill & Radial velocity jitter ($\mathrm{m\,s}^{-1}$)\dotfill & \DRPjitter & \SERVALjitter & - \\
~~~~$v_{\rm macro}$\dotfill & Macroturbulence of the host star ($\mathrm{km\,s}^{-1}$)\dotfill & - & - & \DTnonrotv\\
~~~~$u_{0,\tess}$\dotfill & Quadratic limb-darkening coefficient 0\dotfill & \DRPua & \SERVALua & \DTua\\
~~~~$u_{1,\tess}$\dotfill & Quadratic limb-darkening coefficient 1\dotfill & \DRPub & \SERVALub & \DTub\\
\smallskip\\
\hline
\end{tabular}
\tablecomments{Due to the asymmetric and bimodal shapes of the eccentricity and argument of periapse posteriors, instead of reporting their medians and 68\% credible intervals, we report their 68\% highest posterior density intervals.}
\end{table*}

\acknowledgments
Computations for this research were performed on the Pennsylvania State University’s Institute for Computational \& Data Sciences Advanced CyberInfrastructure (ICS-ACI). This content is solely the responsibility of the authors and does not necessarily represent the views of the Institute for CyberScience. The Center for Exoplanets and Habitable Worlds is supported by the Pennsylvania State University and the Eberly College of Science. We gratefully acknowledge support by NASA XRP 80NSSC18K0355 and NASA TESS GO 80NSSC18K1695.

These results are based on observations obtained with NEID on the WIYN 3.5m Telescope at Kitt Peak National Observatory. WIYN is a joint facility of the University of Wisconsin–Madison, Indiana University, NSF's NOIRLab, the Pennsylvania State University, Purdue University, University of California, Irvine, and the University of Missouri. The authors are honored to be permitted to conduct astronomical research on Iolkam Du'ag (Kitt Peak), a mountain with particular significance to the Tohono O'odham.

This work makes use of observations from the LCOGT network.

This work was performed for the Jet Propulsion Laboratory, California Institute of Technology, sponsored by the United States Government under the Prime Contract 80NM0018D0004 between Caltech and NASA.

The research leading to these results has received funding from the ARC grant for Concerted Research Actions, financed by the Wallonia-Brussels Federation.

TRAPPIST is funded by the Belgian Fund for Scientific Research (Fond National de la Recherche Scientifique, FNRS) under the grant PDR T.0120.21, with the participation of the Swiss National Science Fundation (SNF). TRAPPIST-North is a project funded by the University of Liege (Belgium), in collaboration with Cadi Ayyad University of Marrakech (Morocco) MG and EJ are F.R.S.-FNRS Senior Research Associate.

CDD acknowledges support from the NASA Exoplanet Research Program (XRP) under award 80NSSC20K0250.

A.A.B. and I.A.S. acknowledge the support of Ministry of Science and Higher Education of the Russian Federation under the grant 075-15-2020-780 (N13.1902.21.0039).

We acknowledge the use of public TESS data from pipelines at the TESS Science Office and at the TESS Science Processing Operations Center. 

Resources supporting this work were provided by the NASA High-End Computing (HEC) Program through the NASA Advanced Supercomputing (NAS) Division at Ames Research Center for the production of the SPOC data products.

This research made use of $\mathtt{exoplanet}$ \citep{exoplanet:exoplanet, fore21} and its dependencies \citep{exoplanet:agol20, exoplanet:astropy13, exoplanet:astropy18, exoplanet:exoplanet, exoplanet:foremanmackey17, exoplanet:foremanmackey18, exoplanet:kipping13, exoplanet:luger19, exoplanet:pymc3, exoplanet:theano}.

Some/all of the data presented in this paper were obtained from the Mikulski Archive for Space Telescopes (MAST) at the Space Telescope Science Institute. The specific observations analyzed can be accessed via \dataset[10.17909/t9-nmc8-f686]{https://doi.org/10.17909/t9-nmc8-f686}.

\vspace{5mm}
\facilities{\tess, \emph{Gaia}, LCOGT, TRAPPIST-North, PHARON, \alopeke, ShARCS, KELT, TRES, NEID, Exoplanet Archive}

\software{$\mathtt{ArviZ}$ \citep{Kumar2019}, $\mathtt{astroARIADNE}$, $\mathtt{AstroImageJ}$ \citep{Collins:2017}, $\mathtt{astropy}$ \citep{exoplanet:astropy13, exoplanet:astropy18}, $\mathtt{BAFFLES}$ \citep{stan20}, $\mathtt{celerite2}$ \citep{exoplanet:foremanmackey17, exoplanet:foremanmackey18}, $\mathtt{Comove}$ \citep{toff21}, $\mathtt{EXOFASTv2}$ \citep{Eastman:2013, Eastman:2019}, $\mathtt{exoplanet}$ \citep{fore21}, $\mathtt{Jupyter}$ \citep{kluy16}, $\mathtt{Matplotlib}$ \citep{hunt07, droe16}, $\mathtt{NumPy}$ \citep{vand11, harr20}, $\mathtt{pandas}$ \citep{mckinney-proc-scipy-2010}, $\mathtt{PyMC}$ \citep{exoplanet:pymc3}, $\mathtt{SciPy}$ \citep{2020SciPy-NMeth}, $\mathtt{starry}$ \citep{exoplanet:luger19}, $\mathtt{Tapir}$ \citep{Jensen:2013}, $\mathtt{VARTOOLS}$ \citep{hart16}}

\bibliography{toi1268}{}
\bibliographystyle{aasjournal}

\end{document}